\documentclass[conference,10pt]{IEEEtran}
\usepackage{epsfig,rotating,setspace,latexsym,amsmath,epsf,amssymb,amsfonts,bm,theorem,subfigure,epstopdf}
\usepackage{cite,authblk}
\usepackage{url}
\usepackage{bbm}
\usepackage{algorithm,algpseudocode}

\IEEEoverridecommandlockouts

\allowdisplaybreaks
\usepackage{xcolor}
\usepackage{graphicx}

\definecolor{green1}{rgb}{0.2,0.7,0.2}
\definecolor{brown}{rgb}{1,0.5,0.2}
\usepackage[colorinlistoftodos,textwidth=\marginparwidth]{todonotes}

\begin{document}
\title{Gradient Coding with Dynamic Clustering \\for Straggler Mitigation \\\thanks{This work was supported by EC H2020-MSCA-ITN-2015 project SCAVENGE under grant number 675891, and by the European Research Council project BEACON under grant number 677854.}}
 	\author[1]{Baturalp Buyukates}
	\author[2]{Emre Ozfatura}
	\author[1]{Sennur Ulukus}
	\author[2]{Deniz G\"{u}nd\"{u}z}
	\affil[1]{\normalsize Department of Electrical and Computer Engineering, University of Maryland, MD, USA}  
	\affil[2]{\normalsize Department of Electrical and Electronic Engineering, Imperial College London, UK}
\maketitle
\begin{abstract}
In distributed synchronous gradient descent (GD) the main performance bottleneck for the per-iteration completion time is the slowest \textit{straggling} workers. To speed up GD iterations in the presence of stragglers, coded distributed computation techniques are implemented by assigning redundant computations to workers. In this paper, we propose a novel gradient coding (GC) scheme that utilizes dynamic clustering, denoted by GC-DC, to speed up the gradient calculation. Under time-correlated straggling behavior, GC-DC aims at regulating the number of straggling workers in each cluster based on the straggler behavior in the previous iteration. We numerically show that GC-DC provides significant improvements in the average completion time (of each iteration) with no increase in the communication load compared to the original GC scheme.
\end{abstract}

\section{Introduction}

Common to many machine learning problems is the use of gradient descent (GD) methods to optimize the parameters of the model in an iterative fashion. In large scale learning problems with massive datasets, computations required in GD can be distributed across workers to speed up GD iterations. In a typical distributed synchronous GD implementation in the parameter server (PS) framework, workers first compute gradient estimates, also called \textit{partial gradients}, based on their own local datasets. These partial gradients are then aggregated by the PS to update the model. The main performance bottleneck in this distributed synchronous framework is the slowest \textit{straggling} workers. Recently, straggler-tolerant distributed GD schemes have received significant attention, where the main strategy is to assign redundant computations to workers to mitigate the potential delays due to stragglers either together with coded dataset \cite{CC.1,CC.3,CC.4, CC.6, CC.7, CC.8, CC.11, Ozfatura20}, or combined with coded local computations \cite{UCCT.1,UCCT.2,UCCT.3}, or by simply using backup computations \cite{UCUT.5, UCUT.2,UCUT.4}. 


In this paper, we focus on the gradient coding (GC) setup \cite{UCCT.1}, where the dataset is distributed across the workers in an uncoded but redundant manner, and workers return coded computations to the PS. A static clustering technique is introduced in \cite{Ozfatura19}, which entails dividing the workers into smaller clusters and applying the original GC scheme on a cluster level. This technique is shown to improve the average computation time with respect to the original GC scheme. With clustering, unlike in the original GC scheme, the number of tolerated stragglers scales with the number of clusters when the stragglers are uniformly distributed among the clusters. However, this may not be the case in practical scenarios as evident in the measurements taken over Amazon EC2 clusters that indicate a time correlated straggling behavior for the workers \cite{Yang19}. In this case, the advantage of clustering diminishes since the stragglers are not uniform across clusters.

To mitigate this problem and to further improve the performance, in this paper, we introduce a novel GC scheme with dynamic clustering, called GC-DC. GC-DC aims to assign straggling workers to clusters as uniformly as possible at each iteration based on workers' past straggling behavior. The main idea behind GC-DC is to assign more data samples to workers than the actual computation load (per-iteration) to give them certain flexibility in choosing the computations they carry out at each iteration. By doing so, we enable a worker replacement strategy, which helps us dynamically form clusters at each iteration to tolerate stragglers. We numerically show that the proposed GC-DC scheme significantly improves the average per-iteration completion time of the GC framework with no increase in the communication load.

\section{GC with Clustering}

\subsection{GC}
GC is a distributed coded computation technique to perform GD across $K$ workers \cite{UCCT.1}. To tolerate straggling workers, GC assigns redundant computations to each worker. That is, the dataset $\mathcal{D}$ is divided into $K$ non-overlapping equal-size mini-batches, $\mathcal{D}_{1},\ldots,\mathcal{D}_{K}$, and each worker is assigned multiple mini-batches. We denote the set of indices of mini-batches assigned to the $k$th worker with $\mathcal{I}_{k}$. Let $\boldsymbol{g}^{(t)}_{k}$ denote the partial gradient for the parameter vector $\boldsymbol{\theta}_{t}$ evaluated over mini-batch $\mathcal{D}_{k}$ at the $t$th GD iteration, i.e.,
\begin{equation}
\boldsymbol{g}^{(t)}_{k}=\frac{1}{\vert \mathcal{D}_{k} \vert} \sum_{x \in \mathcal{D}_{k}} \nabla l (x, \boldsymbol{\theta}_{t}),
\end{equation}
where $l$ is the application-specific loss function. We note that the \emph{full gradient} computed over the whole dataset is given by $\boldsymbol{g}^{(t)}=\frac{1}{K}\sum^{K}_{k=1}\boldsymbol{g}^{(t)}_{k}$.

If a mini-batch $\mathcal{D}_i$ is in set $\mathcal{I}_{k}$, then the corresponding partial gradient $\boldsymbol{g}^{(t)}_{i}$ is computed by the $k$th worker. \textit{Computation load}, $r$, denotes the number of mini-batches assigned to each worker, i.e., $\vert\mathcal{I}_{k}\vert = r$, $\forall k\in[K]$. At each iteration, each worker computes $r$ partial gradients, one for each mini-batch available locally, and sends a linear combination of the results, $\boldsymbol{c}^{(t)}_{k}\triangleq\mathcal{L}_{k}(\boldsymbol{g}^{(t)}_{i}:i\in\mathcal{I}_{k})$, called the {\em coded partial gradient}. Thus, in the GC scheme, each worker is responsible for computing a single coded partial gradient. The underlying code structure in GC, which dictates the linear combinations formed by each worker, exploits the available redundancy so that the PS can recover the full gradient from only a subset of the combinations. Accordingly, from now on, we refer to coded partial gradients simply as \emph{codewords.} As shown in \cite{UCCT.1}, the GC scheme can tolerate up to $r-1$ persistent stragglers at each iteration. Formally, for any set of non-straggling workers $\mathcal{W}\subseteq[K]$ with  $ \lvert \mathcal{W}   \rvert = K-r+1 $, there exists a set of coefficients $\mathcal{A}_{\mathcal{W}}=\left \{a^{(t)}_{k}:k\in\mathcal{W}\right\}$ such that
\begin{align}
\sum_{k\in\mathcal{W}} a^{(t)}_{k}\boldsymbol{c}^{(t)}_{k}=\frac{1}{K}\sum^{K}_{k=1}\boldsymbol{g}^{(t)}_{k}.\label{GC_decod}
\end{align}
Thus, at each iteration $t$, the full gradient $\boldsymbol{g}^{(t)}$ can be recovered from any $K-r+1$ codewords.

Next, we present \textit{clustering} \cite{Ozfatura19} that is used to reduce the average per-iteration completion time of the GC scheme.

\subsection{Clustering}

We divide workers into $P$ equal-size disjoint clusters. Let $\mathcal{K}_{p} \subset [K]$ denote the set of workers in cluster $p$, $p\in[P]$, where $\mathcal{K}_q \cap \mathcal{K}_p = \emptyset$ for $q \neq p$. We denote the cluster size by $\ell \triangleq \frac{K}{P}$, where we assume $P|K$ for simplicity. Let $\mathcal{I}_{\mathcal{K}_{p}}$ denote the set of mini-batches assigned to the $p$th cluster. The GC scheme is applied to each cluster separately and the workers in cluster $p$ enable computing
\begin{equation}
\frac{1}{\vert\mathcal{I}_{\mathcal{K}_{p}}\vert}\sum_{k: \mathcal{D}_k \in \mathcal{I}_{\mathcal{K}_{p}} }\boldsymbol{g}^{(t)}_{k}.
\end{equation}

To illustrate the advantage of the clustering technique, consider $K=12$, $r=2$, and $P=4$. Here, the workers are divided into $4$ clusters of size $\ell=3$, and each cluster is responsible for computing $3$ of the total $12$ partial gradients. Since $r=2$, each worker computes $2$ partial gradients. The assignment of the workers to the clusters is dictated by the following matrix, where each column corresponds to a different cluster.
\begin{align}
\mathbf{A}_{cluster}=
  \begin{bmatrix}
{1} & {2} & {3} & {4}\\
{6} & {7} & {8} & {5}\\
{9} & {10} & {11} & {12}\\
  \end{bmatrix}.\label{A_cluster_st}
\end{align}
The corresponding data assignment matrix for static clustering, $\mathbf{A}_{data}$, is given in (\ref{assign_mat_cl}), where the clusters are represented by different colors. In $\mathbf{A}_{data}$, each column $i$ represents the partial gradient computations (correspondingly the mini-batches) assigned to the $i$th worker. Equivalently, in (\ref{codeword_mat}), $\mathbf{A}_{code}$ represents the codewords assigned to the workers, where $c_{p,i}$ corresponds to the codeword assigned to the $i$th worker in the $p$th cluster, for $p \in [P]$, $i \in [\ell]$. Codewords corresponding to different clusters are shown in different colors. Each codeword in $\mathbf{A}_{code}$ is a linear combination of $r=2$ partial gradients. For example, $c_{1,1}$ is a linear combination of partial gradients $g_1$ and $g_2$; $c_{1,2}$ is a linear combination of partial gradients $g_2$ and $g_3$, and $c_{1,3}$ is a linear combination of partial gradients $g_3$ and $g_1$. 

 \begin{figure}[t]
	\centering  \includegraphics[width=0.85\columnwidth]{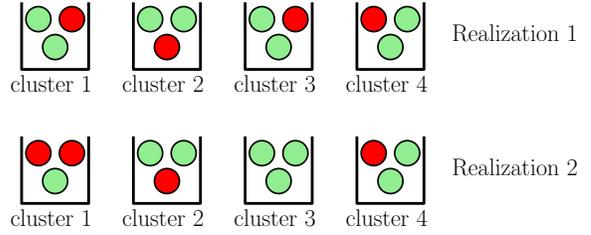}
	\caption{Two possible straggler realizations where red and green circles represent the straggling and non-straggling workers, respectively.}
	\label{fig_cluster}
	\vspace{-5mm}
\end{figure}

\setcounter{MaxMatrixCols}{20}
 \begin{figure*}
 \begin{align}
\mathbf{A}_{data}=
  \begin{bmatrix}
  {\color{blue}g_{1}} & {\color{red}g_{4}} & {\color{magenta}g_{7}} &  {\color{green1}g_{10}} &  {\color{green1}g_{11}} & {\color{blue}g_{2}} & {\color{red}g_{5}} & {\color{magenta}g_{8}}  & {\color{blue}g_{3}} & {\color{red}g_{6}} & {\color{magenta}g_{9}}  & {\color{green1}g_{12}}\\
  {\color{blue}g_{2}} & {\color{red}g_{5}} & {\color{magenta}g_{8}}   & {\color{green1}g_{11}} &  {\color{green1}g_{12}} & {\color{blue}g_{3}} & {\color{red}g_{6}} & {\color{magenta}g_{9}}  & {\color{blue}g_{1}} & {\color{red}g_{4}} & {\color{magenta}g_{7}}  & {\color{green1}g_{10}} \\
  \end{bmatrix}\label{assign_mat_cl} 
  \end{align}
  \vspace{-5mm}
 \end{figure*}
 
In the original GC scheme, the PS waits until it receives $K-r+1=11$ results at each iteration; hence only $1$ straggler can be tolerated. With clustering, the PS needs to receive at least $\ell-r+1=2$ results from each cluster to recover the full gradient. Thus, the non-straggling threshold is still $K-r+1$, since more than one straggler cannot be tolerated if they are in the same cluster. However, the non-straggling threshold represents a worst case scenario. With clustering, up to $4$ stragglers can be tolerated if they are uniformly distributed across clusters, i.e., one straggler per cluster, as shown in ``Realization $1$" in Fig.~\ref{fig_cluster}. This shows that, in the case of clustering, the number of realizations in which the full gradient can be recovered is higher than that of the original GC scheme. Thus, even if the non-straggling threshold remains the same, clustering will reduce the average per-iteration completion time.

Formally, with clustering, it is possible to tolerate $r-1$ stragglers in each cluster in the best case scenario, which is when the stragglers are uniformly distributed among the clusters. In this case, it is possible to tolerate $P(r-1)$ stragglers in total. However, this advantage of clustering diminishes in the case of non-uniform distributed stragglers among the clusters, which may be the case in practice. As shown in ``Realization $2$" in Fig.~\ref{fig_cluster}, even if there are still $8$ non-straggling workers, the PS cannot compute the full gradient (in the case of persistent stragglers) when the stragglers are not uniformly distributed across the clusters. To this end, in the next section, we introduce the concept of dynamic clustering, which dynamically changes the codewords computed by the workers at each iteration based on the past straggler behavior to further improve the performance of the clustering technique.
 
 \setcounter{MaxMatrixCols}{20}
 \begin{figure*}
 \begin{align}
\mathbf{A}_{code}=
  \begin{bmatrix}
{\color{blue}c_{1,1}}&{\color{red}c_{2,1}}&{\color{magenta}c_{3,1}}&{\color{green1}c_{4,1}}&{\color{green1}c_{4,2}}&{\color{blue}c_{1,2}}&{\color{red}c_{2,2}}&{\color{magenta}c_{3,2}}&{\color{blue}c_{1,3}}&{\color{red}c_{2,3}}&{\color{magenta}c_{3,3}}&{\color{green1}c_{4,3}}
  \end{bmatrix}\label{codeword_mat}
  \end{align}
  \vspace{-5mm}
 \end{figure*}
 
\section{GC with Dynamic Clustering (GC-DC)}
In the conventional coded computation approaches, including the GC, the assignment of the dataset to the workers and the code to be used are set at the beginning of the training process. Therefore, in order to recover the desired computation result at each iteration, the codes are designed for the worst case scenario. The core idea behind dynamic clustering is to change the codewords assigned to the workers dynamically based on the observed straggling behavior. Dynamic clustering is driven by two policies; namely, data assignment and codeword assignment. The data assignment policy, denoted by $\Pi_{d}$, is executed only once at the beginning of training and assigns up to $m$ mini-batches to each worker, where $m$ denotes the memory constraint, i.e., 
\begin{equation}
\Pi_{d}:  \mathcal{D} \mapsto \left\{\mathcal{I}_{1},\ldots,\mathcal{I}_{K}: \vert\mathcal{I}_{k} \vert \leq m\right\}.
\end{equation}
We note that even though each worker can be allocated up to $m$ mini-batches, each will compute only $r$ of them at each iteration. Thus, we can have $m \choose r$ codewords that can be assigned to each worker depending on which subset of $r$ computations it carries out among $m$ possibilities. Here, we introduce $\mathcal{C}=\left\{\mathcal{C}_{1},\ldots,\mathcal{C}_{K}\right\}$, where $\mathcal{C}_{k}$ denotes the set of feasible codewords corresponding to dataset $\mathcal{I}_{k}$ and may include codewords corresponding to different clusters.\\
\indent Then, at the beginning of each iteration $t$, codeword assignment policy $\Pi_{c}$ is executed based on the past straggler behavior up to iteration $t$, $\mathbf{S}^{[t-1]}$, i.e.,
\begin{equation}
\Pi^{(t)}_{c}(\mathbf{S}^{[t-1]}, \Pi_d):  \mathcal{C}\mapsto {\mathbf{c}}^{t}= \left\{c^{t}_{1},\ldots,c^{t}_{K}\right\},
\end{equation}
where $c^{t}_{k} \in \mathcal{C}_k$ is the codeword assigned to the $k$th worker at iteration $t$, $\mathbf{S}^{[t-1]} \triangleq (\mathbf{S}^1, \ldots, \mathbf{S}^{t-1})$, and $\mathbf{S}^t=(S^t_1,\ldots, S^t_K)$ denotes the straggler behavior, where $S^t_{k}=0$ if the $k$th worker is a straggler at iteration $t$, and $S^t_{k}=1$ otherwise. We note that, by assigning codewords $\mathbf{c}^t$ to workers, $\Pi_c$ essentially performs dynamic cluster formation at each iteration $t$.  

The completion time of iteration $t$ for a given data assignment policy $\Pi_d$ depends on the codeword assignment ${\mathbf{c}}_{t}$ and the straggler realization $\mathbf{S}^{t}$. Here, our objective is to minimize the expected completion time of each iteration based on the past straggler behavior for a given $\Pi_d$:
\begin{equation}
\min_{\Pi^{(t)}_{c}} \mathbb{E}_{\mathbf{S}^{t}\vert \mathbf{S}^{[t-1]}, \Pi_d} Q(\tilde{\mathcal{C}}_{t},\mathbf{S}^{t}),
\end{equation}
where $Q_{t}(\tilde{\mathcal{C}}_{t},\mathbf{S}^{t})$ is the completion time of iteration $t$ under codeword assignment ${\mathbf{c}}_{t}$ and the straggler realization $\mathbf{S}^{t}$.\\
\indent We remark that the codeword assignment policy $\Pi^{(t)}_{c}$ (correspondingly the dynamic cluster assignment) highly depends on the data assignment policy $\Pi_{d}$ since in most of the coded computation scenarios the data assignment policy is driven by the employed coding strategy. Thus, designing a data assignment policy $\Pi_{d}$ without any prior knowledge on the coding strategy is a challenging task. To this end, in the next section, we reformulate the dynamic clustering problem where the coding strategy, consequently the set of codewords, are fixed at the beginning and data assignment is performed based on the underlying coding strategy.

\section{Solution Approach}
We perform three steps; namely, codeword construction, codeword distribution, and dynamic codeword assignment, where the first two steps are executed once at the beginning of training and the last one is executed at each iteration.

\subsection{Codeword Construction}
We first construct a set of codewords $\mathcal{C}$ according to GC with clustering. Here, the set of codewords $\mathcal{C}$ is a union of smaller disjoint codeword sets, i.e., $\mathcal{C}=\cup^{P}_{p=1}\mathcal{C}^{p}$, such that codewords in each set $\mathcal{C}^{p}$, $p\in [P]$, are encoded and decoded independently and correspond to a particular cluster. For example, in (\ref{codeword_mat}), $\mathcal{C}^{1}=\{c_{1,1}, c_{1,2}, c_{1,3}\}$, where $\mathcal{C}^{1}$ is disjoint from the rest of the codeword set.

\subsection{Codeword Distribution}
Then, the codewords $\mathcal{C}$ are distributed among the workers according to an assignment policy $\Pi_{c}$ i.e.,
\begin{equation}
\Pi_{c}(\mathcal{C}): \mathcal{C} \mapsto \left\{\mathcal{C}_{1},\ldots,\mathcal{C}_{K}\right\},\label{codeword_pol} 
\end{equation}
where we redefine $\mathcal{C}_{k}$ as the set of codewords assigned to the $k$th worker. Now, let $\mathcal{I}(c)\subseteq\mathcal{D}$ be the minimal subset of mini-batches that is sufficient to construct codeword $c$. Given the codeword assignment policy $\Pi_{c}$, any feasible data assignment policy $\Pi_{d}$ should satisfy the following constraint
\begin{equation}
\mathcal{I}_{k}\supseteq\cup_{c\in\mathcal{C}_{k}}\mathcal{I}(c), \quad \forall k \in [K].\label{assignment_pol}
\end{equation}
Based on this constraint we observe that, given $\Pi_{c}(\mathcal{C})$, the minimum memory is used when $\mathcal{I}_{k}=\cup_{c\in\mathcal{C}_{k}}\mathcal{I}(c), \forall k\in [K]$. Thus, we note that the data assignment policy is determined according to the codeword assignment policy. 

Next, we describe the codeword assignment policy $\Pi_{c}$ in (\ref{codeword_pol}). We first assign each worker to $n$ clusters using a circular shift operator where the shift amounts are sampled uniformly at random. Since each cluster $p$ corresponds to a set of codewords $\mathcal{C}^p$ with $|\mathcal{C}^p|=\ell$, this step indicates that each worker is assigned codewords from an $n$-subset of $\{\mathcal{C}^1,\ldots,\mathcal{C}^P\}$. We say that a worker is in cluster $p$, if that worker is assigned all $\ell$ codewords in $\mathcal{C}^p$. With this, we form a worker cluster assignment matrix $\mathbf{A}_{cluster}$ of size $\ell n_{1} \times P$. Here, the $p$th column of $\mathbf{A}_{cluster}$ illustrates the workers assigned to the $p$th cluster, where $w_k$ denotes the $k$th worker, $k\in [K]$. An example $\mathbf{A}_{cluster}$ for $n=2$ is given by
\begin{align}
\mathbf{A}_{cluster}=
  \begin{bmatrix}
w_{1} & w_{2} & w_{3} & w_{4}\\
w_{6} & w_{7} & w_{8} & w_{5}\\
w_{9} & w_{10} & w_{11} & w_{12}\\
w_{4} & w_{1} & w_{2} & w_{3}\\
w_{7} & w_{8} & w_{5} & w_{6}\\
w_{10} & w_{11} & w_{12} & w_{9}\\
  \end{bmatrix}.\label{A_cluster}
\end{align}

We remark that given $n$, the memory requirement $m$ is given by $m=n\ell$.
Thus, for $n=2$ and $\ell=3$, each worker stores $6$ mini-batches, i.e., half of the whole dataset. 

Once $\mathbf{A}_{cluster}$ is constructed, we assign corresponding mini-batches to each worker to form the data assignment matrix such that the constraint in (\ref{assignment_pol}) is satisfied with equality. For example, from (\ref{A_cluster}) we deduce that worker $1$ has all the codewords in sets $\mathcal{C}^1$ and $\mathcal{C}^2$, i.e., $\mathcal{C}_1 = \mathcal{C}^1 \cup \mathcal{C}^2 = \{c_{1,1}, c_{1,2}, c_{1,3}, c_{2,1}, c_{2,2}, c_{2,3}\} $. Correspondingly, $\mathcal{I}_1 = \{\mathcal{D}_1,\ldots,\mathcal{D}_6\}$ so that worker $1$ can compute partial gradients $g_1,\ldots,g_6$ to form any one of these $6$ codewords. 

\subsection{Dynamic Clustering}

The key idea behind dynamic clustering is to assign more than one codeword, and consequently more than $r$ mini-batches, to each worker. That is, each worker is assigned a total of $n\ell$ codewords. We assign one of these codewords to each worker at each iteration based on the previous straggler realization. We note that, even though more than one codeword is assigned to each worker, computation load is still $r$ as in the original GC scheme, since each worker still computes only one codeword consisting of $r$ partial gradients at each iteration. 

This dynamic codeword assignment strategy can be interpreted as a worker replacement scheme, such that each worker can potentially replace the other workers that are assigned to the same cluster with itself, by computing a codeword that would be computed by the worker to be replaced in the original GC scheme with clustering. 

To see the advantage of the dynamic codeword assignment strategy, we consider $\mathbf{A}_{cluster}$ and corresponding codewords for a particular straggler realization $\mathbf{S}=[{\color{blue}1},{\color{red}1},{\color{magenta}0},{\color{green1}1},{\color{green1}1},{\color{blue}1},{\color{red}1},{\color{magenta}0},{\color{blue}0},{\color{red}1},{\color{magenta}0},{\color{green1}1}]$, where, colors follow the cluster assignment in the static clustering case, i.e., $\mathbf{A}_{cluster}$ given in (\ref{A_cluster_st}). In the static clustering case, it is not possible to recover partial gradients corresponding to the third cluster as we do not have $\ell-r+1=2$ non-straggling workers in that cluster. Moreover, if this straggling behavior persists for a substantial duration of time, the overall computation time will suffer drastically. To mitigate this, in the case of dynamic clustering, we observe in (\ref{A_cluster}) that worker $w_5$ can replace worker $w_3$ since it can compute codeword $c_{3,1}$. This does not affect the recoverability of the partial gradients assigned to the fourth cluster, to which worker $w_5$ initially belongs, since that cluster has $2$ more non-straggling workers, workers $w_4$ and $w_{12}$. Further, worker $w_2$ can replace worker $w_8$ so that all partial gradients can be recovered successfully. Equivalently, we have assigned the clusters such that $w_2$ and $w_5$ belong to the $3$rd cluster. Thus, dynamic clustering increases the set of straggler realizations for which the full gradient recovery is possible compared to static clustering. 

Since each worker can replace any worker in all the $n$ clusters that it is assigned to, we essentially form the clusters, dynamically at each iteration through codeword assignments hence the name \emph{dynamic clustering.} Our aim is to dynamically form clusters at each iteration to minimize the average completion time of an iteration given the past straggler behavior and the worker-cluster assignment matrix $\mathbf{A}_{cluster}$. 

In the next section, we propose a greedy worker replacement strategy that aims to uniformly place stragglers across clusters at each iteration to speed up GC.

\section{Greedy Dynamic Clustering Strategy }

In this section, we consider a time-correlated straggler behavior for the workers. At each iteration, the PS identifies the stragglers based on the previous observation and implements a greedy dynamic clustering strategy to uniformly distribute the stragglers across clusters.

Inspired by the bin packing problem \cite{Korte08}, we consider clusters as bins and workers as balls as in Fig.~\ref{fig_cluster}. Unlike the bin packing problem, which aims to place balls of different volumes into a minimum number of bins of finite volume, in our setting, the number of bins (clusters) is fixed and our aim is to distribute the straggling workers as uniformly as possible to clusters. Our dynamic clustering algorithm has two phases: in the first phase, based on the previous straggler realization, we place straggler and non-straggler workers into clusters separately following a specific order, and in the second phase, any placement conflict (i.e., if a worker cannot be placed into any of the remaining clusters) that may happen in the first phase is resolved through worker swap between the corresponding clusters. During worker placement, we implement a greedy policy such that each worker is placed into the first cluster in which it will fit based on the given worker cluster assignment matrix $\mathbf{A}_{cluster}$.

To illustrate the proposed worker replacement policy in detail, we consider the cluster assignment matrix in (\ref{A_cluster}) and without loss of generality, order workers in an increasing order in each column to obtain
\begin{align}
\mathbf{A}_{cluster}=
  \begin{bmatrix}
w_{1} & w_{1} & w_{2} & \color{red}{w_{3}}\\
w_{4} & w_{2} & \color{red}{w_{3}} & w_{4}\\
\color{red}{w_{6}} & \color{red}{w_{7}} & \color{red}{w_{5}} & \color{red}{w_{5}}\\
\color{red}{w_{7}} & \color{red}{w_{8}} & \color{red}{w_{8}} & \color{red}{w_{6}}\\
w_{9} & w_{10} & w_{11} & w_{9}\\
w_{10} & w_{11} & w_{12} & w_{12} \label{ex_cluster}
  \end{bmatrix}.
\end{align}
where straggling workers are shown in red. The straggler realization for this example is $\mathbf{S}=[1,1,0,1,0,0,0,0,1,1,1,1]$. Here, we have $5$ straggling and $7$ non-straggling workers.

\begin{figure}[t]
	\centering  \includegraphics[width=0.95\columnwidth]{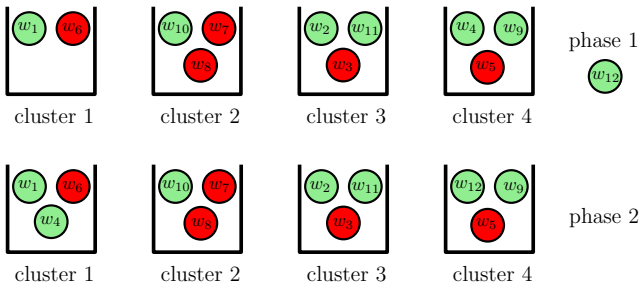}
	\caption{The proposed worker placement strategy.}
	\label{fig_propstrat}
	\vspace{-5mm}
\end{figure}

Since there are more non-straggling workers than stragglers, we place the non-straggling workers first. To determine a non-straggling worker placement order, we find the number of available non-straggling workers in each cluster. One can observe in (\ref{ex_cluster}) that, cluster $1$ and cluster $2$ have $4$ available non-straggling workers that can be assigned to these clusters whereas cluster $3$ and cluster $4$ have $3$ available non-straggling workers. Based on these, we deduce a placement order $O = [3,4,1,2]$ such that clusters take turns based on this placement order.\footnote{In a more refined implementation, this order can dynamically change after each round of worker placement, i.e., after all clusters select one worker, to better reflect the clusters with less availability as worker placement continues.} At each turn of a particular cluster, a single worker is assigned to that cluster according to a greedy policy. In our example, we start with the third cluster and $w_2$ is assigned to this cluster. Then, the fourth cluster gets $w_4$ and so on. This process continues until all the non-straggling workers are placed into clusters (or until a placement conflict is observed). If a cluster is assigned $\ell=3$ workers, we remove that cluster from the order vector $O$. Next, we determine the placement order of straggling workers in a similar fashion. One can deduce from (\ref{ex_cluster}) that the order of placement for the stragglers is $O=[1,2,3,4]$ as clusters $1$ and $2$ have the least availability. Based on this order, stragglers are also placed using a greedy policy and the first phase of the proposed strategy terminates. Worker placement at the end of the first phase for our example is shown in Fig.~\ref{fig_propstrat}. Here, we observe a placement conflict as $w_{12}$ has not been assigned to any cluster whereas cluster $1$ needs one more worker, but $w_{12}$ cannot be assigned there.

With this, we start the second phase of the proposed worker placement algorithm to place $w_{12}$ into a cluster which has a worker that can be assigned to the first cluster. We see from (\ref{ex_cluster}) that $w_{12}$ can be assigned to clusters $3$ or $4$. The algorithm identifies that $w_4$, which has been assigned to the fourth cluster in the first phase can go to the first cluster. With this, we swap workers $w_4$ and $w_{12}$, which yields the final placement in Fig.~\ref{fig_propstrat}.\footnote{We note that at the end of the first phase, there are $4$ other workers, namely workers $w_4, w_7, w_9$, and $w_{10}$, that can be placed into the first cluster. Thus, it is guaranteed that cluster $3$ and cluster $4$ have at least one worker that can be assigned to cluster $1$.} At the end of the algorithm we see that the stragglers are placed into the clusters as uniformly as possible such that cluster $2$ has two stragglers whereas the remaining clusters has only $1$ straggler each. We note that since we have only $7$ non-straggling workers, less than the worst case scenario of $8$ non-stragglers, the full recovery would not be possible for most realizations under a static clustering scheme. Thus, the proposed dynamic clustering with worker replacement does not improve the worst case scenario. Rather, it speeds up the GC scheme by uniformly placing the stragglers across clusters. This process is repeated at each iteration to dynamically change the clusters based on the straggler observations. 

In the next section, we analyze the performance of this dynamic clustering strategy.

\section{Numerical Results}
In this section, we provide numerical results for comparing the proposed GC-DC scheme with GC with static clustering (GC-SC) as well as the original GC scheme using a model-based scenario for computation latencies. For the simulations, we consider a linear regression problem over synthetically created training and test datasets, as in \cite{CC.4}, of size of $2000$ and $400$, respectively. We set the size of the model to $d = 1000$. A single simulation consists of $T=400$ iterations. For all simulations, we use learning rate $\eta = 0.1$. To model the computation delays at the workers, we adopt the model in \cite{entropy}, and assume that the probability of completing $s$ computations at any worker, performing $s$ identical matrix-vector multiplications, by time $t$ is given by
\begin{equation}\label{exp_compute}
  F_s(t) \triangleq
  \begin{cases}
    1-e^{-\mu(\frac{t}{s}-\alpha)}, & \text{if $t \geq s\alpha$} \\
    0, & \text{otherwise}. 
  \end{cases}
\end{equation}
where $\alpha$ is a constant shift indicating that a single computation duration cannot be smaller than $\alpha$.

We model the time-correlated straggling behavior of workers based on a two-state Markov chain: a slow state $s$ and a fast state $f$, such that computations are completed faster when a worker is in state $f$. Specifically, in (\ref{exp_compute}) we have rate $\mu_f$ in state $f$ and rate $\mu_s$ in state $s$ where $\mu_f > \mu_s$ as in \cite{Ozfatura20, Buyukates20a}. We assume that the state transitions only occur at the beginning of each iteration with probability $p$; that is, with probability $1-p$ the state remains the same. A low switching probability $p$ indicates that the straggling behavior tends to remain the same in consecutive iterations with occasional transitions. We set $p=0.05$, $\alpha = 0.01$, $\mu_s = 0.1$, and $\mu_f = 10$.

\begin{figure}[t]
	\centering  \includegraphics[width=0.73\columnwidth]{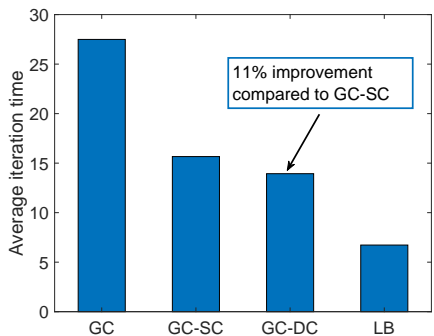}
	\vspace{-1em}
	\caption{Average per-iteration completion time under the shifted-exponential model for $K=12$, $P=4$, $r=2$, $n=2$.}
	\label{sim1}
	\vspace{-5mm}
\end{figure}

In the first simulation we consider a setup with $K=12$ workers and the dataset is divided into $K = 12$ mini-batches. We set $r = 2$; that is, $2$ partial gradient computations, each corresponding to a different mini-batch, are assigned to each worker. We take $P=4$ such that four equal-size clusters are formed. We set $n=2$ and let $6$ of the total $12$ workers start at the slow state, i.e., initially we have $6$ straggling workers. In Fig.~\ref{sim1}, we plot the average per-iteration completion time of the original GC scheme, GC scheme with static clustering, denoted by GC-SC, GC scheme with the proposed dynamic clustering, denoted by GC-DC, and the lower bound, denoted by LB. Here, the lower bound is obtained by assuming that the full gradient is recovered as soon as the earliest $P\times(\ell-r+1)$ workers finish their computations at each iteration, independently of the codeword assignment matrix.

We observe in Fig.~\ref{sim1} that clustering schemes significantly improve the performance compared to the GC scheme. The best performance is achieved when the dynamic clustering is implemented, although the performance improvement with respect to GC-SC is smaller than the performance improvement with respect to plain GC by implementing clustering.

\begin{figure}[t]
	\centering  \includegraphics[width=0.73\columnwidth]{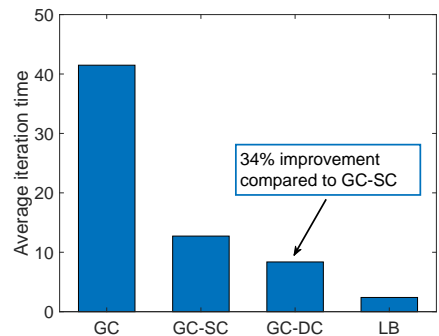}
	\vspace{-1em}
	\caption{Average per-iteration completion time under the shifted-exponential model for $K=20$, $P=5$, $r=3$, $n=3$.}
	\label{sim2}
	\vspace{-3mm}
\end{figure}

In the second simulation, we set $K=20$, $P=5$, $r=3$, and $n=3$. We start with $10$ stragglers initially. In this case, we observe in Fig.~\ref{sim2} that the GC-DC scheme still performs the best and this time the performance improvement compared to the GC-SC scheme ($34\%$) is much more significant. This is due to the increase in the cluster size $\ell$ and the number of assigned clusters $n$, which together increase the dynamic clustering capability of the proposed greedy algorithm.

\begin{figure}[t]
	\centering  \includegraphics[width=0.73\columnwidth]{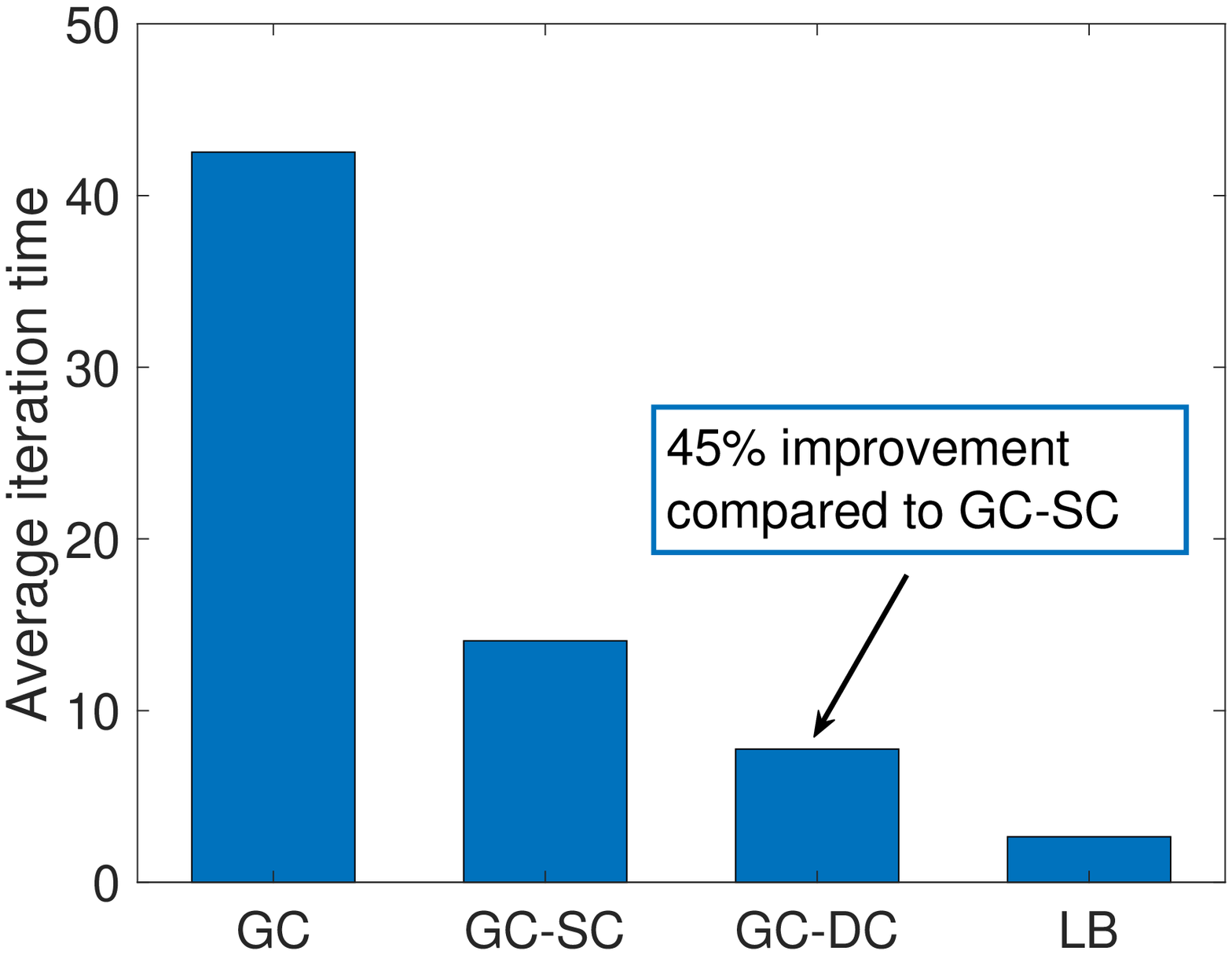}
	\vspace{-1em}
	\caption{Average per-iteration completion time under the shifted-exponential model for $K=20$, $P=5$, $r=3$, $n=3$ under perfect SSI.}
	\label{sim3}
	\vspace{-5mm}
\end{figure}

So far in the simulations, we consider the case in which the PS does not know the exact straggler realization at the beginning of the iteration. Thus, the PS uses previous observation to implement the dynamic clustering. In the third simulation in Fig.~\ref{sim3}, we consider the same setup as in the second simulation but assume that the PS knows the exact straggler realization at the beginning of each iteration, which we coin the perfect straggler state information (SSI). In this case we see similar trends as in Fig.~\ref{sim2}, but observe that the GC-DC scheme results in around $45\%$ improvement in the average per-iteration completion time.

\section{Conclusions}
We considered a gradient coding problem and proposed a greedy dynamic clustering technique, called GC-DC, which assigns additional data to the workers while the computation load per iteration remains the same as in the original GC scheme. Additional data assigned to workers provides extra degree-of-freedom to dynamically assign workers to different clusters by a worker replacement strategy, in order to distribute the stragglers to clusters as uniformly as possible at each iteration. Under a time-correlated straggler model, we showed through numerical simulations that the proposed dynamic clustering technique can drastically improve the average per-iteration completion time with no increase in the communication load. 



\bibliographystyle{unsrt}
\bibliography{IEEEabrv,ref,lib_v5}
\end{document}